\documentstyle[12pt]{article}

\newcommand{\beq}{\begin{equation}}
\newcommand{\eeq}{\end{equation}}
\begin{document}
\title
{A Gauge Invariant Electromagnetic Two-Point Function For Heavy-Light
Quark Systems}

\author{Leonard S. Kisslinger and Zhenping Li\\
      Department of Physics, Carnegie Mellon University\\
      Pittsburgh, PA 15213}

\maketitle
\indent
\begin{abstract}
A gauge invariant electromagnetic two-point function,
crucial to the investigations of the violation of isospin symmetry,
is derived for heavy-light quark systems.  Thus QED can be consistently
introduced into the QCD Sum Rule method, which was not previously possible.
\vspace{3mm} \\

\end{abstract}


The origin of mass differences in isospin multiplets has long been of great
interest in nuclear and particle physics as a source of information about
symmetry violations.
Hadronic isospin violations are particularly important in that they arise from
nonperturbative Quantum Chromodynamics (QCD) as well as quark mass differences
(see Ref.\cite{gl} for a review of the early work in this area), and of course
electromagnetic effects.  Among the first applications of the method of QCD
sum rules was the
study of isospin violations in the $\rho-\omega$ system\cite{svz}, in which it
was recognized that the isospin splitting of the light-quark condensates can
produce effects as large as the current-quark mass splittings and
electromagnetic effects.
Recently the QCD sum rule method has been used to
study the neutron-proton mass difference\cite{hhp,yhhk}, the octet baryon
mass splittings\cite{adi} and the mass differences in the charmed meson
systems (the D and $\rm{D}^*$ scalar and vector mesons)\cite{ei}.

In comparison with hadronic quark models, the QCD sum rule method for
calculating isospin splittings has the advantage that one can directly
use QED field theory rather than rely on models to estimate Coulomb
corrections.  This, however, has not been done.  In our earlier attempt
to calculate the standard two-loop QED contributions to the isospin
mass splittings in heavy-light quark systems [Fig. 1] using the sum rule
method we found\cite{kl} that for the charged mesons,
involving charged currents, the calculation is not gauge invariant.
The objective of the present letter is to develope a gauge invariant
theory for QED phenomena within the QCD sum rule method.  Although our
derivation is for heavy-light quark mesons, the method is quite general
and can be used for baryons as well as mesons.

The basic approach of the QCD sum rule in heavy light quark
systems is to study the two-point function in
the Wilson operator product expansion (OPE), defined by
\begin{eqnarray}\label{i}
\bar \Pi^0_{\mu\nu}(q^2) & = & i\int d^4x e^{iqx} <T(J_{\mu}(x)\bar
J_{\nu}(0))>
\nonumber\\
 & = & \sum_n I^n_{\mu\nu}(q^2) O_n
\end{eqnarray}
for the heavy-light quark current
\begin{equation}\label{ii}
J_{\mu}(x) = {\bar q}(x)\gamma_\mu Q(x).
\end{equation}
The local operators, $\left \{ O_n\right \}$, consisting of quark
and gluon fields and the Wilson coefficients functions, $\left \{
I_{\mu\nu}^n(q^2)\right \}$, have been extensively discussed
in the literature
in the studies of the masses and their decay constants for heavy-light
quark systems.

In order to study the violations of the isospin symmetry,
the electromagnetic effects  should also be written in the framework
of the operator product expansion.
The leading electromagnetic effects in the this approach are the
two-point functions from a two-loop perturbative contribution, whose
Feynman diagrams are shown in Fig. 1.  For the charge neutral
current, one could simply obtain
the two-point functions by changing the gluons in QCD to the photons
in QED in Fig. 1, since the two-point functions has been calculated
in QCD\cite{gb}. Their imaginary parts are
\begin{equation}\label{1}
Im \left [{\bar \Pi}^{p.s}(q^2)\right ]=\frac {3e_q^2M^2}{8\pi^2}(1-x)^2f(x)
\end{equation}
and
\begin{eqnarray}\label{2}
Im \left [{\bar \Pi}^{v}(q^2)\right ]=\frac {e_q^2q^2}{8\pi^2}(1-x)^2
\bigg [ (2+x)(1+f(x))-(3+x)(1-x)
\ln\left (\frac x{1-x}\right )\\ \nonumber
-\frac {2x}{(1-x)^2}\ln(x)-5-2x-\frac {2x}
{1-x}\bigg ]
\end{eqnarray}
for pseudoscalar and vector currents, where $x=\frac {M^2}{q^2}$,
\begin{equation}\label{3}
f(x)=\frac 94+2l(x)+\ln(x)\ln(1-x)+
\left (\frac 52-x-\frac 1{1-x}\right )\ln(x)-
\left (\frac 52-x\right )\ln(1-x)
\end{equation}
and $l(x)=-\int_0^x \ln(1-y)\frac {dy}{y}$ is the Spencer function.
Obviously, this result is gauge invariant.

The calculation of the two-point functions for a charged current
is much more complicated.  Since the Feynman diagrams in Fig. 1
are weighted by the products of different charges, the corresponding
two point functions are not gauge invariant.  The physical origin of
this problem is that under the gauge transformation
\begin{eqnarray}\label{iv}
{\bar q}(x) \to e^{ie_q\Lambda(x)} {\bar q}(x) \nonumber \\
Q(x) \to  e^{ie_Q\Lambda(x)}Q(x) \nonumber \\
A_{\mu}(x)\to A_{\mu}(x)-\partial_{\mu} \Lambda(x),
\end{eqnarray}
we have
\begin{equation}\label{4}
J_{\mu}(x) \to J^\prime_{\mu}(x) =e^{i\Lambda(x)(e_Q+e_q)}{\bar q}(x)
\gamma_{\mu}Q(x)=e^{i\Lambda(x)(e_Q+e_q)}J_{\mu}(x)
\end{equation}
(notice $e_q=-e_{\bar q}$), therefore,  the current $J_{\mu}$
becomes gauge dependent with $e_q+e_Q=e_T$.
Of course, gauge invariance is a fundamental property of
electromagnetic interactions, and any observable obtained using the
current must be gauge invariant.

In order to obtain a solution to the problem, let us start with a
gauge invariant form of the two-point function, which is obtained by
inserting the link operator into the form given in Eq. \ref{i},
with the definition:
\begin{eqnarray}\label{ig}
\Pi_{\mu\nu}(q^2) & = & i\int d^4x e^{iq(x-y)} <T(J_{\mu}(x)e^{iQ_{op}
\int_y^x A_{\alpha}(y)dy^{\alpha}}\bar J_{\nu}(y))>, \nonumber \\
 & = & i\int d^4x e^{iq(x-y)} <T(J^t_{\mu}(x)\bar J^t_{\nu}(y))>
\end{eqnarray}
where $Q_{op}$ is the charge operator and
\begin{equation}\label{jt}
J^t_{\mu}(x) = {\bar q}(x)  e^{iQ_{op}\int_0^x A_{\alpha}(y)dy^{\alpha}}
\gamma_{\mu} Q(x).
\end{equation}
The gauge transformation of the electromagnetic field $A_{\alpha}(x)$
cancels those of quark fields ${\bar q}(x)$ and $Q(x)$ so that
two-point function is gauge invariant.  This is QED analog to the  QCD
treatment introduced\cite{cz} for the pion wave function.

Expanding to order $\alpha_e$ one finds that
there are two currents at the two loop level:
\begin{equation}\label{6}
J_{\mu}^t(x)=J_{\mu}^0(x)+J_{\mu}^e(x).
\end{equation}
The current $J^0_{\mu}(x)$ is given by insertion of photon vertices in
the quark lines of the current given by Eq. \ref{ii}; and the evaluation
of the resulting two-point functions is carried out by
the standard two-loop diagrams shown
in Fig. 1. The charges involved are $e_Q$ and $e_q$ for these processes.
The second current of Eq. \ref{6},
\begin{equation}\label{7}
J_{\mu}^{e}(x) = i e_T {\bar q}(x)
\gamma_{\mu}\int_0^x A_{\alpha}(y)dy^{\alpha}Q(x),
\end{equation}
corresponds to an additional vertex function with the charge $e_T$; and in
the evaluation of the two-point function gives rise to the additional
diagrams shown in Fig. 2, which we now discuss in detail.
The resulting gauge invariant two point function is
\begin{equation}\label{8}
\Pi_{\mu\nu}(q^2)
 =  \Pi_{\mu\nu}^0(q^2)+\Pi_{\mu\nu}^e(q^2) ,
\end{equation}
where
\begin{equation}\label{8i}
\Pi^0_{\mu\nu}(q^2)=i\int d^4x e^{iqx}<T\left ( J_{\mu}^0(x)
\bar J^0_{\nu}(0) \right )>
\end{equation}
at the two-loop level is given by the Feynman diagrams in Fig. 1, and
\begin{equation}\label{8ii}
\Pi_{\mu\nu}^e=i\int d^4x e^{iqx}T\left [
J^0_{\mu}(x)\bar J^e_{\nu}(0)+J^e_{\mu}(x)\bar J^0_{\nu}(0)+J^e_{\mu}(x)
\bar J^e_{\nu}(0)\right ]
\end{equation}
generates additional Feynman diagrams shown in Fig. 2.  These additional
terms introduced by $\Pi_{\mu\nu}^e$ arise from the additional vertex
function corresponding to $J^e_{\nu}$.  The fact that such additional
vertex functions must be present for charged currents for gauge invariance
has been observed previously.  See, e.g., Ref.\cite{gh}.

Since the masses for the heavy and light quarks are not the same,
it is more convenient to calculate the two-point function in
momentum space.  The Fourier transformation
of the operator $\int_0^x A_{\mu}(y)dy^{\mu}$ gives
\begin{eqnarray}\label{9}
F(q^2) & = & \int d^4x e^{iqx} \int_0^x A_{\mu}(y)dy^{\mu} \nonumber \\
& = & \frac {q_\nu}{iq^2} \int d^4x \left (\partial^\nu e^{iqx}\right )
\int_0^x A_{\mu}(y)dy^{\mu} \nonumber \\
& = & -\frac {q_\nu}{iq^2} A^{\nu}(q^2),
\end{eqnarray}
where $q_\mu$ is the momentum carried by photon field $A_{\mu}(q^2)$,
therefore the current $J_{\mu}^e(q^2)$ in the momentum space is given by
\begin{equation}\label{10}
J_{\mu}^e(q^2)=-e_T{\bar q}(q-k_1) \gamma_{\mu} \frac {(k_1-k_2)_{\nu}}
{(k_1-k_2)^2} A^{\nu}(k_1-k_2) Q(k_2).
\end{equation}
The sum of the Feynman diagrams in Fig. 2 has a simple form
\begin{eqnarray}\label{13}
\Pi_{\mu\nu}^e(q^2) & = & -\frac {e_T(e_q+e_Q)}{2^8 \pi^4}
\int \frac {d^D k_1 d^D k_2}{\pi^D} \frac {Tr \left [
 \gamma_\mu \not\! k_2\gamma_\nu (\not\! k_1-\not\! q) \right ]}
{(k_1-k_2)^4(k_2^2-M^2)(k_1-q)^2 } \nonumber \\
& = & e_T(e_q+e_Q) I_{\mu\nu}(q^2).
\end{eqnarray}

To show that the addition of $\Pi^e_{\mu\nu}(q^2)$
indeed makes the total two-point function, $\Pi_{\mu\nu}(q^2)$,
gauge invariant,  we separate $\Pi^0_{\mu\nu}(q^2)$ into the gauge
dependent and independent parts;
\begin{eqnarray}\label{14}
\Pi^0_{\mu\nu}(q^2) & = & e_q^2 I^q_{\mu\nu}(q^2)+e_Q^2 I^Q_{\mu\nu}(q^2)
-e_qe_Q I^{qQ}_{\mu\nu}(q^2) \nonumber \\
& = & -e_qe_Q\left (I^q_{\mu\nu}(q^2)+I^Q_{\mu\nu}(q^2)+I^{qQ}_{\mu\nu}
(q^2) \right ) \nonumber \\ & &
+e_Te_q I^q_{\mu\nu}(q^2)+e_Te_Q I^Q_{\mu\nu}(q^2),
\end{eqnarray}
where $I^q_{\mu\nu}(q^2)$ and $I^Q_{\mu\nu}(q^2)$ represent the
the self-energy diagram for the light quark ${\bar q}(x)$ and heavy
quark $Q(x)$ respectively, and $I^{qQ}_{\mu\nu}(q^2)$ corresponds to
the photon exchange between the light and heavy quarks in Fig. 1.
The term with $I^q_{\mu\nu}(q^2)+I^Q_{\mu\nu}(q^2)+
I^{qQ}_{\mu\nu}(q^2)$ in Eq. \ref{14} is of course
gauge invariant, and the analytical expressions of its imaginary
part can be obtained from Eqs. \ref{1} and \ref{2}.
The last two terms in Eq.
\ref{14} are gauge dependent; if we substitute the gauge dependent
part of the photon propagator $\left (D_{\mu\nu}\right )_{gd}
=\frac {(k_1-k_2)_\mu
(k_1-k_2)_\nu}{(k_1-k_2)^4}$ into the two-loop integrals
$I^q_{\mu\nu}(q^2)$ and $I^Q_{\mu\nu}(q^2)$, the gauge
dependent part of the integrals is
\begin{equation}\label{15}
\left (I^{q}_{\mu\nu}(q^2)\right )_{gd}=\left (I^{Q}_{\mu\nu}(q^2)
\right )_{gd}=-I_{\mu\nu}(q^2)
\end{equation}
where $I_{\mu\nu}(q^2)$ is given in Eq. \ref{13}.  By substituting the
gauge dependent photon propagator into $\Pi^e_{\mu\nu}(q^2)$
one can show that the gauge dependent part of the two-loop
integral for $\Pi^e_{\mu\nu}(q^2)$ is identical to the expression in
Eq. \ref{13}.
Therefore, the gauge dependent two-loop integrals
in $\Pi^0_{\mu\nu}(q^2)$ and $\Pi^e_{\mu\nu}(q^2)$ exactly
cancel each other, and we have a total gauge invariant results.
This provides an important check  at the two-loop level
that the two point function $\Pi_{\mu\nu}(q^2)$ for the
current $J^t_{\mu}(x)$ in Eq. \ref{6}
is indeed gauge invariant.

To evaluate the two point function $\Pi_{\mu\nu}(q^2)$,
we rewrite Eq. \ref{8} as
\begin{equation}\label{16}
\Pi_{\mu\nu}(q^2)=\Pi_{\mu\nu}^{qQ}(q^2)+\Pi_{\mu\nu}^q(q^2)
+\Pi_{\mu\nu}^Q(q^2),
\end{equation}
where
\begin{equation}\label{16i}
\Pi_{\mu\nu}^{qQ}(q^2)
=-e_qe_Q\left (I^q_{\mu\nu}(q^2)+I^Q_{\mu\nu}(q^2)+
I^{qQ}_{\mu\nu}(q^2) \right )
\end{equation}
whose analytical expressions can be obtained from Eqs. \ref{1}
and \ref{2},
\begin{equation}\label{16ii}
\Pi_{\mu\nu}^q(q^2)
=e_Te_q \left (I^q_{\mu\nu}(q^2)+I_{\mu\nu}(q^2)\right )
\end{equation}
and
\begin{equation}\label{16iii}
\Pi^Q_{\mu\nu}(q^2)=e_Te_Q \left (I^Q_{\mu\nu}(q^2) +I_{\mu\nu}(q^2)\right ).
\end{equation}
For $\Pi^q_{\mu\nu}(q^2)$, we have
\begin{equation}\label{17}
I^q_{\mu\nu}(q^2)+I_{\mu\nu}(q^2)=\frac 1{2^8\pi^4} \int
\frac {d^D k_2}{\pi^{D/2}}
 \frac {Tr \left [ \gamma_\mu (\not\! k_2+M) \gamma_\nu (\not\! k_2-\not\! q)
F(k_2) (\not\! k_2-\not\! q) \right ]}
{(k_2^2-M^2)(k_2-q)^4 } ,
\end{equation}
where
\begin{equation}\label{18}
F(k_2)=\int \frac {d^D k_1}{\pi^{D/2}} \frac {\gamma_{\mu}
(\not\! k_1-\not\! q)\gamma_{\nu}}{(k_1-k_2)^2(k_1-q)^2} \left (
g^{\mu\nu}-\frac {(k_1-k_2)^{\mu}(k_1-k_2)^{\nu}}{(k_1-k_2)^2}\right )
\end{equation}
is an one-loop wavefunction renormalization  equivalently
in Landau gauge.  It has been shown that $F(k_2)$ vanishes\cite{muta}
for a zero mass particle in dimensional regularization.  This leads to
\begin{equation}\label{19}
\Pi_{\mu\nu}^Q(q^2)  = e_Te_Q(I^Q_{\mu\nu}(q^2)-I^q_{\mu\nu}(q^2))
\end{equation}
where $\Pi^Q_{\mu\nu}(q^2)$ is proportional to the mass of the heavy
quarks.  This shows that the divergence induced by the wavefunction
renormalizations does not exist in Eq. \ref{19}, which implies that
the Ward identity is restored in this approach.

After including the mass renormalization, the two-loop integral
 $\Pi^Q_{\mu\nu}(q^2)$ in the dimensional regularization is
\begin{equation}\label{21}
\left (\frac {\mu^2 e^\gamma}{4\pi}\right )^{\epsilon}
\frac {\Pi^Q_{\mu\nu}(q^2)}{e_Qe_T}=3 \left (\frac {\mu^2}
{-q^2}\right )^\epsilon \left [ -\frac {3\alpha_e}{4\pi
\epsilon} {\bar M}\frac {\partial}
{\partial {\bar M}} I^0_{\mu\nu}(q^2)+\left (\frac {\mu^2}
{-q^2}\right )^\epsilon \left (I^Q_{\mu\nu}(q^2)-I^q_{\mu\nu}(q^2)
\right)\right ],
\end{equation}
where
\begin{equation}\label{22}
I^0_{\mu\nu}(q^2)=-i\frac {(-q^2e^\gamma)^\epsilon}{16\pi^2}
\int \frac {d^D k}{\pi^{D/2}} \frac {Tr \left ( \gamma_\mu (\not\! k+M)
\gamma_\nu (\not\! k-\not\! q)\right ) }{(k^2-M^2)(k-q)^2}
\end{equation}
is the one-loop integral, $D=4-2\epsilon$, $\epsilon \to 0^+$, is the
number of spacetime dimensions, and
the running mass ${\bar M}(\mu)$ is related to the
pole mass $M$ by\cite{gb}
\begin{equation}\label{23}
{\bar M}(\mu)=M\left \{ 1-\frac {\alpha_e}{4\pi} \left [
3\ln \left (\frac {\mu^2}{M^2}\right )+4\right ]\right \}.
\end{equation}
The evaluation of Eq. \ref{21} is
performed in the modified minimal subtraction scheme ($\overline
{\mbox{MS}}$).
 Here we only present
the imaginary parts of $\Pi^1_{\mu\nu}(q^2)$ for the pseudo-scalar
and the vector states;
\begin{eqnarray}\label{23}
Im \left (\Pi^Q_{p.s}(q^2)\right )=\frac {3\alpha_e e_Qe_TM^2}{16\pi^2}
\bigg [ 2x(1-x)+2\ln(x)+(1-x)^2\ln \left (\frac {1-x}x\right )
\nonumber \\ +\frac 32
(1-x)^2\bigg ]
\end{eqnarray}
and
\begin{eqnarray}\label{24}
Im \left (\Pi^Q_{v}(q^2)\right )=\frac {9\alpha_e e_Qe_T}{16\pi^2}
\bigg [ q^2(2+x)\left (x(1-x)+\ln(x)+(1-x)^2\ln \left (\frac {1-x}x\right )
\right ) \nonumber \\
+M^2\left (x(1-x)+\ln(x)+\frac 32
(1-x)^2\right )\bigg ]
\end{eqnarray}
for pseudo-scalar and vector currents respectively, in which the
running mass ${\bar M}(\mu)$ is replaced by the pole mass $M$.
In the limit of $M\to 0$, $\Pi^Q(q^2)$ vanishes for both
pseudo-scalar and vector currents.  Thus,
the two point functions are dominated by the term proportional to $e_qe_Q$
in the small $M$ limit,  this is exactly what one would expect
from the phenomenological model.  However, how large effects the
contributions from $\Pi^1(q^2)$ are in the heavy quark limit remains
 to be studied.

In summary, we have derived a gauge-independent form for QED corrections to
QCD sum rules by identifying processes with additional vertices which must be
included for charged currents.
The result here presents an important step toward a
consistent treatment of the violations of isospin symmetry in the
QCD sum rule framework.  The applications of this result to isospin splittings
of heavy-light quark systems as well as the Kaon systems are
in progress, and will be given elsewhere.

This work is supported by National Science Foundation grant
PHY-9023586.

\vspace{5mm}
\noindent {\bf Figure Captions}

\vspace{5mm}
\begin{itemize}
\begin{enumerate}
\item The two loop perturbative corrections for charged
neutral currents
\item Additional diagrams generated by $J^e_{\mu}(x)$ for charged
heavy-light quark systems. See text.
\end{enumerate}
\end{itemize}
\end{document}